\documentclass[showpacs,preprintnumbers,amsmath,amssymb,prl,twocolumn]{revtex4}
\usepackage{graphicx}
\usepackage{dcolumn}
\usepackage{bm}
\def\comment#1{}

\def\xxi{\xi}

\begin{document}

\title{Photoproduction in  semiconductors
by onset of magnetic field}

\author{Hagen Kleinert}
\email{kleinert@physik.fu-berlin.de}
\affiliation{Institut f{\"u}r Theoretische Physik, Freie Universit\"at Berlin, 14195 Berlin, Germany}
\author{She-Sheng Xue}
\email{xue@icra.it}
\affiliation{ICRANet, Piazzale della Repubblica, 10 - 65122 Pescara, and
Physics Department, University of Rome ``La Sapienza", Italy}

\date{Received \today}

\begin{abstract}

The energy bands of
a semiconductor are lowered
by an external
magnetic field.
When a field is switched on,
the straight-line trajectories near the top of
the
occupied valence
band are curved into Landau orbits and Bremsstrahlung
is emitted until the electrons have settled in their final Fermi distribution.
We calculate the radiated energy, which
should be experimentally detectable, and suggest
that a semiconductor can be cooled
by an oscillating  magnetic field.
\end{abstract}

\pacs{71.55.Cn,
71.70.Ej,42.50.Xa }

\maketitle

\vspace{-2cm}

\section{Introduction}\label{int}

Valence and conduction electrons
near the
Fermi sphere
of a semiconductor have many similarities
with
the Dirac electrons in the vacuum.
In fact, the hole state
as a missing state in the valence band
is completely analogous to a positron in Dirac's sea
of occupied negative-energy
electrons.
The band width $ \Delta $ of a superconductor,
 which is typically of the order of
$0.1$eV,
corresponds to the energy gap $ \Delta =2m_ec^2\simeq 1.04\,$MeV
in Dirac's vacuum, above which electron-positron pairs can be produced.
As a consequence,
the electromagnetic behavior of a semiconductor
at and below room temperature with
$k_BT\simeq 0.024$eV
 ($k_B$=Boltzmann constant)
can be studied
by the same field-theoretic techniques as a Dirac vacuum
for
$k_BT\ll 2mc^2$.
In particular, one can transfer
the results
found by Heisenberg and Euler \cite{HE1936,DUNNE}
for electrons and positron
to electrons and holes.
A strong electric field larger than $E_c=m_e^2c^3/e\hbar \simeq 1.3\cdot 10^{18}\, {\rm V/m}$ leads to
electron-hole pair production.

A magnetic field $H$ lowers the energy of the ground state
since the electrons are curved into Landau orbits \cite{landau}.
This
should produce
synchrotron radiation. For magnetic field
switched on in the vacuum,
this was pointed out in
ref. \cite{xue} as a consequence of the Euler-Heisenberg
calculation
 \cite{HE1936,DUNNE}.
However, this effect could become observable
only for extremely large magnetic fields
which cannot be attained in present-day laboratories.
It will play a role
mainly in astrophysical events, such as
supernova explosions,
and during the formation of
neutron stars, where
magnetic fields
reach $H_c=m_e^2c^3/e\hbar= 4.3\times 10^{13}$Gauss.
It may also
account for the emission of an anomalous X-ray pulsar \cite{xray}.

The
purpose of this note is to suggest
observing this type of synchrotron radiation
at presently available magnetic fields
of $10^5$ Gauss by placing a semiconductor in a magnetic field.
Moreover, we point out that
this may give rise to a novel cooling
technique for semiconductors.

\section{Electron and Hole States}

The electrons in the highest valence band
of a semiconductor
occupy
Bloch states which look like free-particle states
$\psi_{{\bf k}}({\bf r},t)=e^{i{\bf k}{\bf x}-i{\xxi}_{{\bf k}}t }
\psi_{{\bf k}}({\bf r})$,
where ${\bf k}$
is the Bloch momentum and  ${\xxi}_{{\bf k}}$ the energy measured from the
Fermi surface between the bands.
Near the top of the band, the energy can be expanded as
\cite{solidbook}
\begin{equation}
{\xxi}_{\bf k}\simeq  \Delta _v+
 \sum_i\frac{k^2_i}{2m^*_i},  ~~~ (i=x,y,z),
\label{spectrum1}
\end{equation}
where $ \Delta _v$ is the distance of the top of the
valence band from the Fermi level, which is usually close to $ \Delta /2$,
and
$m_i$ are the effective masses
in the three space directions.
The constant-energy surface ${\xxi}_{\bf k}={\rm const.}$ is in general
an ellipsoid.
For simplicity, we shall assume axial symmetry with $m^*_x=m^*_y\equiv m^*_\perp$, and
shall switch on the magnetic field in the $z$-direction.
Then the energies
(\ref{spectrum1})
are replaced by the Landau energies
\begin{equation}
{\xxi}(n,k_z)=\Delta_v+ \left[\frac{k^2_z }{2m_z^*}
+(n+
 \lambda
)\omega^*\right],
\quad \omega^*=\frac{eH}{m_\perp^*},
\label{spectrum2}
\end{equation}
where the constant $\lambda$ is independent of $n$, $k_z$ and $H$ \cite{solidbook}.
\comment{
$ \sigma ^*\equiv
{1}/{2}+ \sigma g^*  $, $\sigma=\pm $ and $g^*$  is the gyromagnetic ratio
of the magnetic moment of the electron inside the solid.
}
The area of the Landau orbits $A({\xxi}_{\bf k},k_z)$
is quantized,
\begin{equation}
A(n,k_z)=(n+ \lambda )\Delta A,\quad \Delta A
\equiv {2\pi eH},
\label{quantum}
\end{equation}
where $n=0,1,2,\cdot\cdot\cdot$. Eqs. (\ref{spectrum2},\ref{quantum}) are Onsager's famous result.

Given a semiconductor sample of volume $V=L_x L_y L_z$,
the electrons of a fixed quantum number $n$ occupy a
phase space $L_xL_y\Delta A$. Dividing this by
the volume $(2\pi)^2$ per quantum state,
we obtains the degeneracy of the states of fixed $n$:
\begin{equation}
{\mathcal D}=2\frac{L_xL_y\Delta A}{(2\pi)^2}=2 \frac{L_xL_y eH}{2\pi},
\label{degeneracy}
\end{equation}
where the factor 2 accounts for the spin degeneracy and ${\mathcal D}$ is independent of $n$.
This is of course the same as
for
free electrons.
\begin{equation}
\frac{2 e}{2\pi\hbar}\simeq 4.84\cdot 10^6 \frac{1}{{\rm cm}^2{\rm Gauss}},
\label{degeneracy1}
\end{equation}
the degeneracy
for
a $1\times 1$ cm${}^2$ sample
 in a  field of one  kiloGauss
is about $10^{10}$.

\section{Energy Difference}

Summing the energy spectrum (\ref{spectrum2}) over all states in phase space,
we obtain the energy of semiconductor sample,
\begin{equation}
E^H_{\rm tot}
=2\frac{V\Delta A}{(2\pi)^2}\sum_{n}
\int\frac{dk_z}{2\pi}{\xxi}(n,k_z).
\label{energyh}
\end{equation}
From this we have to subtract the energy at $H=0$:
\begin{eqnarray}
E_{\rm tot}=2V\!\int\!\frac{dk_x dk_ydk_z}{(2\pi)^3}
\!\left[\Delta_v+
\left(\frac{k^2_\perp}{2 m^*_\perp}+\frac{k^2_z}{2m_z^*}\right)\right].
\label{energy0}
\end{eqnarray}
where $k^2_\perp=k^2_x+k^2_y$.
To subtract this from
(\ref{energyh}), we express $k^2_\perp$
with the help of a
continuous number $
n={k^2_\perp}/{2 m^*_\perp \omega^*}$
 as
$\int dk_x dk_y=\Delta A\int_0^\infty dn$,
and rewrite (\ref{energy0}) as
\begin{eqnarray}
E_{\rm tot}= 2\frac{V\Delta A}{(2\pi)^2}\!\int_0^\infty \!dn\!\int\!\frac{dk_z}{(2\pi)}
{\xxi}(n,k_z).
\label{energy}
\end{eqnarray}
The energies have ultraviolet divergencies
at large $k_z$,
which we regularize
with a smooth cutoff function $f(k_z)$ equal to unity
for small $|k_z|\ll \Lambda_z$ and vanishing for large $|k_z|\gg \Lambda_z$.
The cutoff $\Lambda_z$ is roughly equal to $\pi/a$ where $a$ is the lattice spacing.
In this way, we obtain
convergent $k_z$-integrals
\begin{equation}
F(n)\equiv \! \int_0^\infty\!\frac{dk_z}{2\pi}f(k_z)
{\xxi}(n,k_z),
\label{energyn}
\end{equation}
and a finite difference between
the energies (\ref{energyh}) and (\ref{energy})
\begin{eqnarray}
\Delta E \equiv  E^H_{\rm tot}-E_{\rm tot}
=4\frac{V\Delta A}{(2\pi)^2}\left[ \sum_{n}F(n) -\int_0^\infty \!dn F(n) \right].\nonumber
\end{eqnarray}
We can now use the Euler-MacLaurin formula and the Bernoulli numbers $B_2=1/6,B_4=-1/30,\cdot\cdot\cdot$
to obtain
\begin{eqnarray} \!\!\!\!\!\!
\Delta E
=4\frac{V\Delta A}{(2\pi)^2}\left[
-\frac{1}{2!}B_2F'(0)-\frac{1}{4!}B_4F^{'''}(0)+\cdot\cdot\cdot\right],
\label{emeq}
\end{eqnarray}
where
\begin{equation}
F'(0)=\omega^*\!\int_0^\infty\!\frac{dk_z}{2\pi}f(k_z),
~~ F^{'''}(0)=0,~~\dots\,,
\label{df}
\end{equation}
so that
\begin{equation}
\Delta E =-V\alpha H^2 \frac{2}{3m^*_\perp}\!\int_0^\infty\!\frac{dk_z}{2\pi}f(k_z).
\label{delta2}
\end{equation}
Choosing $f(z)=\exp (-k_z^2/ \Lambda _z^2)$, we find
\begin{equation}
\Delta E =-V\alpha H^2 \frac{ \Lambda _z}{m^*_\perp}\frac{1}{6\pi^{1/2}},
\label{dekz}
\end{equation}
In principle, we must sum over all
different energy bands in eq. (\ref{delta2}), but only the
bands above and below the Fermi surface will contribute.

The energy difference (\ref{delta2})
is negative. This has several consequences:\\
(i) The semiconductor acts
as a paramagnetic medium with permeability
\begin{equation}
\mu=
1-{\alpha \over6\pi} \frac{ \Lambda _z}{m^*_\perp}
\approx1-{\alpha \over6\pi} \frac{ \pi}{a m^*_\perp}.
\label{@}\end{equation}
This should be checked
by experiment.\\
(ii) When turning on the magnetic field,
the energy difference should be released by the semiconductor.
This can proceed by phonon and photon production.
The first will heat the system, the second could in principle be detected
if there is enough surface.
In the following section, we compute rate and spectrum
of spontaneous photon emission.

\section{Spontaneous photon emission}

Let us turn the
magnetic field $H(t)$ adiabatically on over a long time interval
$ \Delta t=t^+-t^-$:
\begin{equation}
H(t)=\left\{ {0, ~~ t= t^-\rightarrow-\infty\atop H, ~~ t= t^+\rightarrow+\infty}\right.,
\label{bt}
\end{equation}
so that
the time variation of $H(t)$ is
much slower than the time scale of Bloch electron
states in the semiconductor, which is characterized by the period
$\tau _\perp=2\pi/\omega^*$ of the motion in the
$xy$-plane, and by the time scale $\tau _z=\hbar/(m_z^*c^2)$
in the $\hat {\bf z}$ direction.
The normalized initial and final Bloch states
are
\begin{eqnarray}
\psi_i &=&{1\over V^{1/2}}\exp\{i[k_xx+k_yy\bar k_zz-{\xxi}_{\bf k}t]\},\nonumber\\
\psi_f &=& {\chi(y)\over (L_xL_z)^{1/2}}\exp\{i[k'_xx+\bar k'_zz-
{\xxi}_n(k'_z)t]\},
\label{final}
\end{eqnarray}
where the normalized function $\chi(y)$ are \cite{landau},
\begin{eqnarray}
\chi(y)&=& N_\chi^{1/2}e^{-\xi^2/2}H_n(\xi),
\quad
N_\chi=\frac{(eH)^{1/2}}{2^n n!\pi^{1/2}},\nonumber\\
\xi &=& \left(eH\right)^{1/2}\left[y-\frac{k'_x}{eH}\right],
\label{chi}
\end{eqnarray}
and $H_n(\xi)$ is the Hermite polynomial.
The energies of these
states
 are
(\ref{spectrum1}) and (\ref{spectrum2}), respectively.
Due to
axial symmetry w.r.t. $z$-direction,
we start form an initial state with $k_y=0$,
which will remain zero, as seen most easily form
the
semiclassical equation of motion:
${d{\bf k}}/{d t} = ({e}/{m}){\bf k}\times {\bf H}$.

The probability amplitude
for spontaneous photon emission
is \cite{LandauQuantumElectrodynamics}
\begin{equation}
J_{fi}(q)\!=\!-\!ie\!\int dtd^3x \hskip0.05cm \psi^*_f\psi_i \left(\frac{1}{2\omega_qV}\right)^{1/2}\!
\exp\{i(\omega_qt-{\bf q}\cdot{\bf x})\},
\label{j1}
\end{equation}
where $(\omega_q=|{\bf q}|,{\bf q})$
is the photon energy-momentum, and
the photon field is normalized to
one photon energy $\omega_q$ crossing
a unit area per unit time.
The integral
over $t,x,z$
 gives rise to $\delta$-functions for energy and $x,z$ momentum
conservations.
The integral over $y$ is done using the formula \{eq. (7.376) in \cite{gr}\}
\begin{eqnarray}
\int dye^{-iq_yy}e^{-\xi^2/2}H_n(\xi)&=& (-i)^n\left({2\pi \over eH}\right)^{1/2}
{\mathcal A}\nonumber\\
&\times  & H_n(\beta q_y),
\label{j}
\end{eqnarray}
where $\beta^2\equiv 1/eH$ and ${\mathcal A}\equiv \exp\{-iq_y\left({k'_x/eH}\right)-\beta^2q_y^2/2\}$. As a result, eq. (\ref{j1}) becomes
\begin{eqnarray}
J_{fi}(q)&=& (-i)^{n+1}e\left(\frac{N_\chi}{2\omega_qVL_yL_x^2}\right)^{1/2}
({2\pi \over eH})^{1/2}\nonumber\\
&\times & {\mathcal A}H_n(\beta q_y)\label{jr}\\
&\times &(2\pi)^2\delta[{\xxi}_n(k'_z)-{\xxi}_{\bf k}+\omega_q ]
\delta(k_x-k'_x-q_x),\nonumber
\end{eqnarray}
where $q_z=0$ and $\omega_q=(q_x^2+q_y^2)^{1/2}$ for $k_z=k_z'$.
The photons are emitted perpendicular to the $\hat{\bf z}$-direction,
as in synchrotron radiation.
The squared transition amplitude is
\begin{eqnarray}
|J_{fi}(q)|^2 &\!=\!& e^2\left(\frac{N_\chi}{2\omega_q VL_yL_x}\right)
{2\pi \over eH}\nonumber\\
&\!\times\!&H^2_n(\beta q_y)\Delta t \exp (-\beta^2q_y^2)  \label{re0}\\
&\!\times\!&(2\pi)^2\delta[{\xxi}_n(k'_z)\!-\!{\xxi}_{\bf k}\!+\!\omega_q ]
\delta(k_x\!-\!k'_x\!-\!q_x).\nonumber
\end{eqnarray}
This has to be summed over all
final states  to yield
summing $n$ with degeneracy ${\mathcal D}$ (\ref{degeneracy}),
and obtain,
\begin{eqnarray}
&&\!\!\!\!\!\!\sum_f |J_{fi}(q)|^2 = e^2\left(\frac{1}{\omega_q V}\right)
\left(|e|H\right)^{1/2}\Delta t \exp (-\beta^2q_y^2)\nonumber\\
&&~~\times \sum_{n=0}^\infty{1\over2^n\pi^{1/2}n!}H^2_n(\beta q_y)\label{re}
 (2\pi)
\delta[{\xxi}_n(k_z)-{\xxi}_{\bf k}+\omega_q ].~~~~~~~
\end{eqnarray}
The $\delta$-function for the total energy-conservation can be rewritten as
\begin{eqnarray}
&&\!\!\!\!\!\!\!\!\!\!\!(2\pi)\delta[{\xxi}_n(k_z)\!-\!{\xxi}_{\bf k}+\omega_q ]
\nonumber\\&& \!\! \!\!\!\!\!\!
= (2\pi)\delta[(n+\sigma^* )\omega^* \!-\! \frac{k_x^2}{2m^*_\perp}\! - \!\omega_q ]
={(\omega^*)^{-1}}\delta_{n,n_{k_x}},
\label{delta}
\end{eqnarray}
where  $n_{k_x}\geq1$ is the integer closest to
$ {(\omega^*)^{-1}} \left( {k_x^2}/{2m^*_\perp} + \omega_q  \right)-
\sigma^*$.
From  (\ref{re}) we obtain the probability of spontaneous photon emission
from the semiconductor per unit time
\begin{equation}
\frac{dN_\gamma}{dt} = {e^2\over 2^{n_{k_x}}\pi^{1/2} n_{k_x} !}
\left(\frac{eH}{\omega_q^2}\right)^{1/2}
H^2_{n_{k_x}}(\beta q_y)\exp (-\beta^2q_y^2),
\label{re1}
\end{equation}
where we can replace $q_y$ by $ \omega _q$ since $q_x=0$.
Multiplying (\ref{re1}) by $ \omega _q$ yields the emitted energy flux.

Note that
the emitted photon energies $\omega_q$
are mostly
smaller than the magnetic energy scale $(eH)^{1/2}$.
For large $n_{k_x}$-values,
the rate (\ref{re1}) has the power-like suppression
$\left[{\omega_q}/{(eH)^{1/2}}\right ]^{2n_{k_x}}/{n^{k_x} !}\ll 1$.
The leading contribution to energy production
comes from the
electrons with $n_{k_x}=1$
which yield (recalling that $H_1(x)=2x$),
\begin{eqnarray}
\frac{d E_\gamma}{dt} &\!\simeq \!& {2e^2\over\pi^{1/2}}
\left(\frac{eH}{\omega_q^2}\right)^{\frac{1}{2}}\!
\exp\left(\!-\!\frac{\omega_q^2}{eH}\right)\frac{\omega^3_q}{eH},
\label{re3e}
\end{eqnarray}
showing a power behavior
 $\propto\omega_q^2$
in the low-energy and an exponential
falloff  $\exp[-\omega_q^2/(eH)]$
at high photon energies.
The spectrum is maximum at $\omega_q=\omega_q^{\rm max}=(eH)^{1/2}$.
Today's laboratories reach
$H=10^5$ Gauss for which
the $\sqrt{eH}\simeq 0.244$eV
\cite{magnet}.
Thus, in present magnetic fields,
most photons are infrared.

Integrating
(\ref{re3e}) over all photon energies,
$\int d\omega_q/(2\pi c)$,  we obtain
the total energy
flux
\begin{eqnarray}
\frac{dE_\gamma}{dt} &\simeq &\alpha
\,eH.
\label{snp1}
\end{eqnarray}

There are two degenerate valence band maxima, both located at ${\bf k}=0$, in
 Silicon as well as Germanium.
In the quadratic approximation
(\ref{spectrum1}), these
 are spherically symmetric, {\it i.e.} $m^*_\perp$ is assumed to be
close to $m_z^*$.
 The effective masses $m^*_\perp$ are $0.49m_e$
and $0.1m_e$ in Silicon, and
$0.28m_e$ and $0.044m_e$ in Germanium \cite{solidbook}.
For a rough estimate,
we take $m^*_\perp=0.1m_ec^2$ and $\sqrt{eH\hbar c}\simeq 0.244$eV for $H=10^5$Gauss.
The characteristic time scales are
$t^*=2\pi/\omega^*\simeq 3.3\cdot 10^{-7}$sec
and $\hbar/(m^*_\perp c^2)=1.3\cdot 10^{-20}$ sec, so that rate
(\ref{snp1}) is of the order of
\begin{equation}
\frac{dE_\gamma}{dt} \simeq 6.6\cdot 10^{11}{\rm eV}{\rm sec}^{-1}.
\label{estimation}
\end{equation}
This should be observable by infrared light detectors
placed around the sample orthogonal to the direction of the
magnetic field $H\approx10^{5}$Gauss.
rates (\ref{estimation}).

\section{Remarks}

In general,
the rate and spectrum of spontaneous photon emission
will
depend on the time dependence
of the magnetic field $H(t)$.
Our
calculation yields only a first estimate
based
 on the adiabatic condition
$\Delta t\gg t^*=2\pi/\omega^*=3.3\cdot 10^{-7}$sec, which
is fulfilled in most experiments.
On these time scales, one need not worry
about  photons generated by induction.
These have a typical energy $\hbar/\Delta\tau$,
which is much smaller than the typical energy $\sqrt{eH\hbar c}$ of
the photons in (\ref{estimation}).
Of course, the experiments should be performed at low temperature to
reduce the background of thermal photons,
which is about $10^8/{\rm cm}^3$ at the room temperature.

\comment{
All these discussions and calculations in the present article can be also applied to the energy spectrum of
metals at low temperature. At low temperature, the Fermi surfaces of energy spectrum of metals very sharp,
{\it i.e.} very few electrons are exited on the Fermi surface. The skin and screening effect can be avoided and
the magnetic field can penetrate whole volume of metals. The effect of spontaneous photon emission
should be produced by the same mechanism as in the semiconductor case.
}

For a periodically oscillating magnetic field $H(t)=H_0\cos(\omega_H t)$
with a period $
t_H=2\pi/\omega_H\gg t^*$,
we can still apply the adiabatic approximation. In the phase that the magnetic field increases from $0$ to maximum $H_0$,
the spontaneous photon emission occur as discussed. If these photons
 are not kept inside the sample by either a large opacity or
reflection on the walls of a cavity, they
will stream away and carry off energy.
In the phase where the magnetic field decreases from the maximum $H_0$ to $0$,
 the
semiconductor
may absorb heat from the environment.
As a consequence, a periodically
 oscillating magnetic
field $H(t)$
should be able to cool a semiconductor.
If the sample is sufficiently thin
the electron phonon coupling should not produce enough
phonons to destroy the effect by re-heating.
Hence there is a good chance
that this process may have technical applications.

\end{document}